\documentclass[preprint,12pt]{elsarticle}

\usepackage{graphicx}
\usepackage{psfrag}
\usepackage{amssymb}

\journal{Journal of Computational Physics}

\begin{document}

\begin{frontmatter}


\title{Symmetries without symmetries in Smoothed Particle Hydrodynamics}

\author{Juan P. Cruz}
\ead{dirak3d@ifm.umich.mx}
\author{Jos{\'e} A. Gonz\'alez}
\ead{gonzalez@ifm.umich.mx}
\address{Instituto de F\'{\i}sica y Matem\'aticas,
        Universidad Michoacana de San Nicol\'as de Hidalgo. Edificio C-3,
        Cd. Universitaria,
        C. P. 58040 Morelia, Michoac\'an, M\'exico.}

\begin{abstract}
We introduce a technique to solve numerically the relativistic Euler's 
equations in scenarios with spherical symmetry using the standard 
Smoothed Particles Hydrodynamics method in cartesian coordinates. 
This implementation allow us to increase the resolution of the simulations
in order to obtain accurate results. We test our implementation studying the 
evolution of a perfect fluid in a blast wave configuration in a fixed space-time . 
The technique can be easily generalized to axial symmetric problems.
\end{abstract}

\begin{keyword}
Hydrodynamics \sep SPH \sep Numerical Implementation

\end{keyword}

\end{frontmatter}

\section{Introduction}
\label{sec:Intro}
Physical scenarios involving fluids are studied using different
numerical methods. One of these standard methods is the Smoothed 
Particle Hydrodynamics (SPH).

Different implementations for Newtonian and relativistic Euler's equations 
in three spatial dimension using SPH have been studied 
for many years \cite{Monaghan83,Barnes,Gadget,Rosswog,Abel}. 
The idea is to write the evolution equations in cartesian 
coordinates and using a Lagrangian scheme we follow the evolution
of the elements of the fluid during the simulation.
When the physical problem has spherical symmetry,
the standard approach is to rewrite the evolution equations in 
spherical coordinates, try to find the best way to adjust the
parameters of the discretization and then evolve the system under 
that symmetry \cite{Omang}.

In this article we use the ideas introduced in \cite{Alcubierre} 
(in the context of numerical evolutions of black holes) 
in order to evolve the system of equations written in cartesian 
coordinates without rewriting the system of equations, only
using the symmetries of the problem. 
This ``Cartoon SPH'' technique, can easily be generalized 
to systems with axial symmetry. It is very helpful and 
straightforward task if a standard SPH code is already working, 
providing a simple way to obtain high resolution to evolve the 
systems with symmetries and obtain accurate results.

The structure of the paper is the following: In section 
\ref{sec:standard_SPH} we describe the standard SPH. 
In section \ref{sec:hyd_rel_equa} we describe relativistic Euler's
equations. Then, in section \ref{sec:cartoon_SPH} 
we describe the idea and the implementation of the cartoon SPH. 
In section \ref{sec:tests} we present tests for the cartoon
SPH implementation and finally, in section \ref{sec:conclusions} 
we conclude.

\section{Standard SPH}
\label{sec:standard_SPH}

The Smoothed Particle Hydrodynamics is a method used to solve numerically 
hydrodynamical equations. It is a mesh free method, also called a 
Lagrangian method, because we are not dealing with a fixed grid, instead 
we use several nodes called particles distributed on the volume of the 
fluid that we are studying.

The discretization of the functions and their derivatives in the SPH
method, is carried out in two steps \cite{Liu,Monaghan83}:

\begin{enumerate}
\item Integral representation of a function: Let $f$ be a real 
valued function from $R^3 \to R$. We use the following identity:

\begin{equation}\label{eq:Integral_representation}
f(\mbox{\boldmath{r}})=\int_{R^3} \delta(\mbox{\boldmath$r$}-\mbox
{\boldmath$r$}') d\mbox{\boldmath$r$}' \approx \int_\Omega f(\mbox
{\boldmath$r$}) W(\mbox{\boldmath$r$}-\mbox{\boldmath$r$}',h) d \mbox
{\boldmath$r$}'
\end{equation}

where the delta function has been approximated by the function $W$ 
called the kernel and $h$ -called the smoothing length of the kernel-
defines the region where the kernel is different from zero, i.e. 
$\Omega \subset R^3$. The kernel is a smooth function over $R^3$ 
specifically over $\Omega$ and it is normalized to the unity according 
to $\int_\Omega W d \mbox{\boldmath$v$} =1$. $W$ is assumed 
to be symmetric, i.e. it only depends on the norm of the vector 
$\mbox{\boldmath$r$}-\mbox{\boldmath$r$}'$.

\item Particle approximation: We change the integration by a sum 
over discrete volume elements  $d\mbox{\boldmath$r$} \to \Delta V= 1/n(
\mbox{\boldmath$r$})$ where $n$ is the number density, subdividing the 
fluid in $N$ parts we get

\begin{equation}\label{eq:Particle_approx}
\int_\Omega f(\mbox{\boldmath$r$}) W(\mbox{\boldmath$r$}-\mbox{
\boldmath$r$}',h) d \mbox{\boldmath$r$}' \approx 
\sum^N_{b=1} \frac{f_b}{n_b} W_{ab} := <f>_a,
\end{equation}
where we use the convention that for any real valued function 
$f_a:=f(\mbox{\boldmath$r$}_a)$, with $a=1,\dots,N$.
\end{enumerate}

The derivatives of any real valued function are obtained using
the compact support 
	
\begin{eqnarray}
\mbox{\boldmath$\nabla$} f(\mbox{\boldmath$r$}) &=& 
\int_\Omega  \mbox{\boldmath$\nabla$} f(\mbox{\boldmath$r$}) W(\|\mbox{
\boldmath$r$}-\mbox{\boldmath$r$}'\|) d \mbox{\boldmath$r$}' \nonumber\\
&=& \int_\Omega 
f(\mbox{\boldmath$r$}') \mbox{\boldmath$\nabla$} W(\|\mbox{\boldmath$r$}-
\mbox{\boldmath$r$}'\|) d \mbox{\boldmath$r$}' 
+ \int_{\partial \Omega} 
f(\mbox{\boldmath$r$})' W(\|\mbox{\boldmath$r$}-\mbox{\boldmath$r$}'\|) 
\mbox{\boldmath$n$} ds \,\,\, .
\end{eqnarray} 

The integral over the boundary of the volume $\partial \Omega$ 
is zero because of the compact support of the kernel. Then
	
\begin{equation}
<\mbox{\boldmath$\nabla$} f>_a= \sum_b \frac{f_b}{n_b} 
\mbox{\boldmath$\nabla$} W_{ab}.
\end{equation}
	
Following the same procedure we obtain the approximation for the 
divergence of a vector:
 
 \begin{equation}
<\partial_i f^i>_a= \sum_b \frac{1}{n_b} \mbox{
\boldmath$f$}_b \cdot \mbox{\boldmath$\nabla$}_a W_{ab}.
 \end{equation}

\section{Hydrodynamic Relativistic Equation}
\label{sec:hyd_rel_equa}

If we want to study the behavior of a fluid in a curved space time
we need to use the laws of thermodynamics in curved-space-times, i.e.,
local baryon conservation, the first and second laws of thermodynamics
plus the local law of energy-momentum conservation 
\cite{Wheeler}:

\begin{equation}
\label{eq:local_em_conservation}
\mbox{\boldmath$\nabla$} \cdot \mbox{\boldmath$T$} =0.
\end{equation}

Choosing a coordinate basis $\{ \partial_\mu \}$, we express 
equation (\ref{eq:local_em_conservation}) as
$\nabla_\mu T^{\mu \nu}=0$.

Now, we  assume the fluid can be approximated by a perfect fluid
represented by the following stress-energy tensor:
$T^{\mu \nu}=(\rho w + q) u^\mu u^\nu+ g^{\mu \nu}(p+q) $.
Here $\rho$ is the rest mass-energy density, $w=1+ \epsilon + p/\rho$ is 
the relativistic specific enthalpy, $\epsilon$ is specific internal energy, 
$p$ is the hydrodynamic pressure and $q$ is a quantity known as the 
artificial viscosity \cite{SieglerRiffert, Wheeler}.

In order to evolve the system for a Lagrangian formulation of relativistic 
hydrodynamic equations we need to do a $3+1$ splitting of the
space-time. The standard way to do this is using the ADM formalism,
where the space-time is decomposed into an infinite foliation of
spatial hyper-surfaces $\Sigma_t$ of constant $t$ coordinate.
The line element is given by
\begin{equation}
  ds^2= g_{\mu \nu} dx^\mu dx^\nu= -(\alpha^2- \beta^i \beta_i)dt^2 + 2 
\beta_i dx^i dt + \eta_{ij}dx^i dx^j \,\,\, ,
\end{equation}
where $\alpha$ is the lapse function, $\beta^i$ the shift vector, and 
$\eta_{ij}$ the induced 3-metric on $\Sigma_t$ \cite{AlcubBook, Wald, 
JamesGrant}. Greek indices run from $0$ to $3$ and Latin indices from
$1$ to $3$.

We can express the quantities either in the coordinate basis 
$\left\{ \partial_\mu \right\}$ or in the basis formed by the 
Eulerian observer $4-$velocity  $\mbox{\boldmath$n$}$ 
and the spatial vector basis 
$\partial_i$, $\left\{ \mbox{\boldmath$n$}, \partial_i \right\}$
(with $\mbox{\boldmath$n$}$ and $\partial_i$ orthogonal).
For example, we can express the 4-velocity of a fluid in these two basis: 
$\mbox{\boldmath$u$}=\gamma ( \mbox{\boldmath$n$}+ \bar{v}^i \partial_i)$ 
or $\mbox{\boldmath$u$}=\frac{\gamma}{\alpha}(\partial_t+v^i 
\partial_i)$ where $v^i= \alpha \bar{v}^i- \beta^i$ and $\gamma$ is
the Lorentz factor.

In order to obtain the Lagrangian equations of relativistic hydrodynamics 
we define the Lagrangian derivative as
\begin{equation}\label{eq:lagrangian_derivative}
  \frac{d}{dt}= \frac{\alpha}{\gamma} u^\mu \partial_\mu=\partial_t + v^i 
\partial_i \,\, ,
\end{equation}
and we write the equations of motion in terms of this derivative.

The local conservation of baryon number is given by
\begin{equation}\label{eq:baryon_number_conservation}
  \frac{dD^*}{dt} + D^* \partial_i v^i=0,
\end{equation}
where 
\begin{equation}\label{eq:restmassdensity}
D^*=\sqrt{-g} \gamma \rho/\alpha= \sqrt{\eta} \gamma \rho,
\end{equation}
is the relativistic rest mass density, and $\sqrt{\eta}=\sqrt{-g}/\alpha$ 
is the determinant of the induced $3-$metric.
 
With the spatial part ($\mu=i$) of equation (\ref{eq:local_em_conservation}) 
we obtain the relativistic momentum equation 
\begin{equation}
\label{lagmomentum}
\frac{d}{dt}S_i = -\frac{1}{D^*}\left[\partial_i\left[\sqrt{-g}(p+q)\right] -
\frac{\sqrt{-g}}{2}T^{\alpha\beta}\partial_i g_{\alpha\beta}\right] \enspace ,
\end{equation}
with the relativistic specific momentum $S_i$ defined by
\begin{equation}
\label{momentumvar}
S_i = \left(w+\frac{q}{\rho}\right)\gamma\eta_{ij}\bar v^j \enspace .
\end{equation}
Notice that equation (\ref{lagmomentum}) contains
spatial derivatives of the metric of the
space-time. For simplicity, we assume that the fluid
interacts very weakly with the space-time, and we are going to use a 
fixed background. This means that the metric is given during all our
simulation and we can compute the spatial (and also the temporal)
derivatives either analytically or numerically.

The temporal part ($\mu=0$) of equation
(\ref{eq:local_em_conservation}) give us the relativistic energy equation:

\begin{equation}\label{lagenergy}
\frac{d\bar{E}}{dt} =
-\frac{1}{D^*}\left[\partial_i\left[\sqrt{-g}(p + q)v^i\right] +
\frac{\sqrt{-g}}{2}T^{\alpha\beta}\partial_tg_{\alpha\beta}\right] \enspace ,
\end{equation}

with $\bar{E}=\alpha E- \beta^i S_i$ and $E$ is the total relativistic 
specific energy

\begin{equation}\label{energyvar}
E = \left(w + \frac{q}{\rho}\right)\gamma - \frac{p + q}{\rho\gamma} \enspace .
\end{equation}

Finally the system of equations must be closed with an equation of 
state $p=p(\rho,\epsilon)$, we are going to use an ideal gas equation 
of state

\begin{equation}\label{eq:eos}
p=(\Gamma-1) \rho \epsilon.
\end{equation}

\subsection{Discretization of Motion Equations}

Using the ideas and equations introduced in section
\ref{sec:standard_SPH},  the equation of the relativistic momentum 
can be written as:

\begin{eqnarray}
\label{sphmomentumeq}
\frac{d}{dt}\mbox{\boldmath$S$}_a & = &-\sqrt{-g_a} \sum_b
m_b\left(\frac{p_a+q_{ab}}{D^{*2}_a} +
\frac{p_b+q_{ba}}{D^{*2}_b}\right)\mbox{\boldmath$\nabla$}_aW_{ab}\nonumber\\
&&{}-\frac{\sqrt{-g_a}}{D^*_a}\left[(p_a+q_a)\mbox{\boldmath$\nabla$}_a
\left(\ln{\sqrt{-g}}\right)_a-\frac{1}{2}T^{\alpha\beta}_a
\mbox{\boldmath$\nabla$}_a\left(g_{\alpha\beta}\right)_a\right]
\enspace ,
\end{eqnarray}

where $\mbox{\boldmath$S$}_a={\{S_i\}}_a$, and the metric gradients
$\mbox{\boldmath$\nabla$}\ln\sqrt{-g}$ and
$\mbox{\boldmath$\nabla$}g_{\alpha\beta}$ can be calculated from the
given metric.

The relativistic energy equation is
\begin{eqnarray}
\label{sphenergyeq}
\frac{d\bar{E}_a}{dt} & = &
-\frac{\sqrt{-g_a}}{2} \sum_b m_b\left(\frac{p_a+q_{ab}}{D^{*2}_a} +
\frac{p_b+q_{ba}}{D^{*2}_b}\right) (\mbox{\boldmath$v$}_a +
\mbox{\boldmath$v$}_b) \mbox{\boldmath$\cdot\nabla$}_a W_{ab} \nonumber\\
&&{}-\frac{\sqrt{-g_a}}{D^*_a} \left[ (p_a+q_a) \mbox{\boldmath$v$}_a 
{\mbox{\boldmath$\cdot\nabla$}_a \left(\ln{\sqrt{-g}}\right)}_a + \frac{1}{2}
T^{\alpha\beta}_a {\left(g_{\alpha\beta,t}\right)}_a \right]
\enspace .
\end{eqnarray}

Finally, there are two possible ways to obtain the density: 
The first one is recovering the density by summation, using equation 
(\ref{eq:Particle_approx})
\begin{equation}
\label{sphdensity}
D^*_a = \sum_b m_b W_{ab} \enspace ,
\end{equation}
where we have used $n_a=D^*_a/m_a$. 

The second one is integrating the density using equation 
(\ref{eq:baryon_number_conservation}) 

\begin{equation}
\label{sphconti}
\frac{d}{dt}D^*_a = - \sum_b m_b(\mbox{\boldmath$v$}_b -
\mbox{\boldmath$v$}_a)\mbox{\boldmath$\cdot\nabla$}_aW_{ab} \enspace .
\end{equation}

It can be proved that keeping $h$ constant, these two equations are equivalent.

\subsection{Artificial Viscosity}

	In order to handle the shocks that appear evolving Euler equations, we use the extra term
	  $q$ as we mentioned before. This term is the artificial viscosity and it is
	  inspired in the standard artificial viscosity used in \cite{Viscosity1, Viscosity2, SieglerRiffert}.
	
In our simulations, we have used the following artificial viscous pressure
\begin{equation}
	q_a=\frac{1}{2} \Sigma_b m_b \left( \frac{q_{ab}}{D^*_a}+ \frac{q_{ba}}{D^*_b}\right)W_{ab}
\end{equation}
where
\begin{equation}
\label{artvisc}
q_{ab} = \left\{\begin{array}{ll} \rho_a w_a \left[ - \tilde\alpha c_a h_a
{\left( \mbox{\boldmath$\nabla\cdot v$} \right)}_a + \tilde\beta h_a^2
{\left( \mbox{\boldmath$\nabla\cdot v$} \right)}_a^2 \right] &
\mbox{if}\enspace {\left( \mbox{\boldmath$\nabla\cdot v$} \right)}_a < 0
\\ 0 & \mbox{otherwise}\end{array}\right..
\end{equation}
The divergence of the velocity for the a$-th$ particle is 
\begin{equation}
\label{sphdivv}
{\left( \mbox{\boldmath$\nabla\cdot v$} \right)}_a \approx
\frac{\mbox{\boldmath$v$}_{ab} \mbox{\boldmath$\cdot r$}_{ab}}
{|\mbox{\boldmath$r$}_{ab}|^2 + \tilde\epsilon \bar h^2_{ab}} \enspace ,
\end{equation}
where $c_a=\sqrt{\Gamma p_a/(\rho_aw_a)}$ is the relativistic sound velocity
measured in the rest frame of the fluid, $\tilde\alpha$, $\tilde\beta$ and
$\tilde\epsilon$ are numerical parameters, $\mbox{\boldmath$v$}_{ab}$,
$\mbox{\boldmath$r$}_{ab}$ were define as the differences
$\mbox{\boldmath$v$}_{ab} = \mbox{\boldmath$v$}_a - \mbox{\boldmath$v$}_b$,
$\mbox{\boldmath$r$}_{ab} = \mbox{\boldmath$r$}_a - \mbox{\boldmath$r$}_b$,
and $\bar h_{ab}=(h_a + h_b)/2$ is the mean value of the smoothing lengths of
particles $a$ and $b$. This version of artificial viscosity \cite{SieglerRiffert} is equivalent to the invented by Monaghan et. al in \cite{Monaghan83}.

It is also possible to use other methods that offer to solve numerically the discontinuities using a solution of the Riemman problem for the fluid motion equation  \cite{Inutsuka, Molteni, Cha}. That kind of analysis will be part of future works.

\subsection{About the Implementation}

At the initial time $t_0$ we have the initial data of our physical problem 
$(\mbox{\boldmath$r$},\mbox{\boldmath$v$},p,\rho,\epsilon)_{t_0}$.
With this information, we reconstruct the initial relativistic
variables $(D^*,\mbox{\boldmath$S$}$,$\bar{E})_{t_0}$ using 
equations (\ref{eq:restmassdensity}, \ref{momentumvar}, \ref{energyvar}). 
We can now integrate the evolution equations to obtain the
relativistic variables at the new time 
$(D^*,\mbox{\boldmath$S$}, \bar{E})_{t_0+\delta t}$.

The next step is to recover the physical variables. This 
can be accomplished solving numerically an algebraic equation for $\gamma$

\begin{equation}
\label{gamzeroeq}
0 = \left(S^2 - \tilde E^2\right) \gamma^4 + 2G\tilde E\gamma^3+\left(\tilde
E^2 - 2GS^2 - G^2\right)\gamma^2 - 2G\tilde E\gamma +
G^2\left(1+S^2\right)
\end{equation} 
where 
\begin{equation}
\label{ssquare}
S^2 = \eta^{ij}S_iS_j = \left(w + \frac{q}{\rho}\right)^2 \left(\gamma^2 -
1\right)
\end{equation}
and
\begin{displaymath}
\tilde E = E + \frac{q}{\Gamma D}\enspace .
\end{displaymath}

Once the value of $\gamma$ is known, it can be calculated the
rest-mass density $\rho$ using equation (\ref{eq:restmassdensity}),
then the thermodynamic pressure $p$ from the equation
\begin{equation}
w+q/\rho=(\tilde{E} \gamma - G)/(\gamma^2-G) 
\end{equation}
($G=1-1/\Gamma$) and the equation of state (\ref{eq:eos}), also we can 
obtain the specific internal energy $\epsilon$ from 
$\epsilon=\frac{p}{(\Gamma-1) \rho}$ and finally the
velocity $\bar v^i$ from equation (\ref{momentumvar}) using $\eta^{ij}S_j=(w
+ q/\rho)\gamma\bar v^i$.

\section{Cartoon SPH}
\label{sec:cartoon_SPH}

Now we are going to describe how the symmetry of the problem can be
used to improve the numerical calculations.  
We are going to describe the method using spherical symmetry. 
The generalization to cylindrical symmetry is straightforward.

Using the cartesian and the spherical coordinate vector basis
$(\hat{i}, \hat{j}, \hat{k})$ and $(\hat{r},\hat{\theta},\hat{\phi})$ 
respectively, we identify the axis $\hat{k}$ with the radial 
direction $\hat{r}$. 
Then, we subdivide the sphere in $N_s$ shells. We impose that
each shell has the same mass with the following relation:

\begin{equation}\label{eq:find_shells_t}
\Delta m_{i,i+1}= 4 \pi \int^{r^s_{i+1}}_{r^s_i} \rho(r) r^2 dr = 
\frac{M_T}{N_s}
\end{equation}

with $M_T$ the total mass of the configuration and $r^s_i$ 
the inner boundary position of shell $i$  ($i=1,\dots,N_s$). 
To assign the position of the {\it real particles} or nodes we use
	
\begin{equation}\label{eq:find_shells}
\int^{r^m_i}_{r^s_i} \rho(r) r^2 dr = \int^{r^s_{i+1}}_{r^m_i} \rho(r) r^2 dr,
\end{equation}

where $r^s_i$ is the position of the $i$-th real particle.
	
The third step consists in generate a set of new particles 
around each real particle. This can be accomplished building an sphere 
of radius $h_i$ ( {\it smoothing length}) and subdividing it in $N_v$ 
{\it virtual particles}. In order to use the 3D SPH code we need to
assign values of the physical quantities for each virtual particle. 
By construction we know the position and volume of the
virtual particles, then, the physical values required $(\mbox{
\boldmath$r$}^v,\mbox{\boldmath$v$}^v,p^v,\rho^v,\epsilon^v)$ can be 
assigned interpolating the values of the real particles. 
We can compute the mass of each one of the virtual particles using 
the simple relation $m^v=\rho^v \Delta V^v$. Finally, we can 
obtain the auxiliar variables $(D^{*v},\mbox{\boldmath$S$}^v,
\bar{E}^v)$ and use the 3D SPH equations of motion to evolve the 
$N_s$ particles using for each one $N_v$ virtual particles. 

\subsection{Constructing the Virtual Particles}
Given a physical particle located at $r^m_i$, we construct a sphere of 
radius $h_i$ containing $N_n$ physical particles inside of it.

We use spherical coordinates and split the $\theta$ and $\phi$ angles 
in $N_\theta$ and $N_\phi$ parts respectively. 
The radial coordinate is subdivided in $N_r+1$ parts. 
The additional subdivision of the radial coordinate corresponds 
to the central physical particle. 

The positions of the boundaries of the volume elements with 
respect to the position of the physical particle, are obtained in
the following way:

\begin{itemize}
\item The radius $h_i$ is subdivided in $N_r+1$ shells, so the first division
is
\begin{equation}
r^b_o= h_i/(N_r+1).
\end{equation}
The rest are obtained demanding that all the shells have the same volume,
using the recurrence relation
\begin{equation}
	r^b_{\alpha}=\left[ r^b_{\alpha-1}+N^{-1}_r h^3_{\alpha-1} \left( 1- (N_r+1)^{-3}\right) \right] \,\,\,\, {\rm with} \,\,\,\,
\alpha=1,\dots,N_r
\end{equation}

\item $\theta$ is subdivided in $N_\theta$ equal parts
  
\begin{equation}
\theta^b_\beta= \beta \cdot \frac{\pi}{N_\theta}, \,\,\,\, {\rm with} \,\,\,\,
\beta=1,\dots, N_\theta \,\,\,\, {\rm and} \,\,\,\, \alpha \ne 1.
\end{equation}

\item $\phi$ is subdivided in $N_\phi$ equal parts

\begin{equation}
\phi^b_\gamma= \gamma  \cdot \frac{2\pi}{N_\phi}, \,\,\,\, {\rm with} \,\,\,\,
\gamma=1,\dots,N_\phi,\,\,\,\, {\rm and} \,\,\,\, \alpha \ne 1.
\end{equation}

\end{itemize}

We identify the position of the virtual particles with the geometrical 
centers of these volume elements:
\begin{eqnarray}
r^v_\alpha &=& \left( \frac{r^b_{\alpha }+r^b_{\alpha-1}}{2} \right), \nonumber \\
\theta^v_\beta &=& \frac{2\beta-1}{2}  \cdot \frac{\pi}{N_\theta}, \\
\phi^v_\gamma&=& \frac{2\gamma-1}{2}  \cdot \frac{2 \pi}{N_\phi}. \nonumber
\end{eqnarray}

The first radius is $r^v_1=\frac{3}{2} \Delta r$ because the
central particle has radius $\Delta r$ and its volume is

\begin{equation}
\Delta V^v_{central}=\frac{4}{3} \pi \Delta r^3
\end{equation}
with $\Delta r = \frac{h_i}{N_r+1}$.

The volume of each element constructed can be calculated by an
analytical expression

\begin{equation}\label{eq:v_v}
\Delta V^v_{\alpha,\beta,\gamma}= -\frac{2 \pi}{3N_\phi}
(\cos(\theta^b_{\beta})-\cos(\theta^b_{\beta-1})) 
\left( r^b_{\alpha}-r^b_{\alpha-1}\right). 
\end{equation}

\subsubsection{Assigning values to the virtual particles}

Now we must assign physical values to each one of the virtual particles 
in order to evolve the system with the SPH algorithm in cartesian coordinates.

The virtual sphere has been subdivided entirely in $N_v=(N_r+1)\cdot N_\theta 
\cdot N_\phi$ volume elements with a virtual particle at the center. 
Lets assign to each one of them a physical value related with the
physical quantities.

Instead of interpolating for the $N_v$ virtual particles, we
use a subset of $N_{fan}=(N_r+1)\cdot N_\theta$ auxiliary particles lying
on $xz$-plane ($\phi^v=0$).

\begin{figure}[ht]
\psfrag{to}{$\theta^b_o=0$}
\psfrag{tm}{$\theta^v_1=\frac{\pi}{2N_\theta}$}
\psfrag{t1}{$\theta^b_1=\frac{\pi}{N_\theta}$}
\psfrag{tj}{$\theta^b_\beta$}
\psfrag{tmmj}{$\theta^v_\beta$}
\psfrag{tjm1}{$\theta^b_{\beta+1}$}
\psfrag{r1}{$r^b_\alpha$}
\psfrag{r2}{$r^v_\alpha$}
\psfrag{r3}{$r^b_{\alpha+1}$}
\psfrag{r_int}{$r^{int}_{\alpha,\beta}$}
\psfrag{rv}{$r^v_\alpha$}
\psfrag{phivzero}{$\phi^v=0$}
\psfrag{x}{$y$}
\psfrag{z}{$x$}
\psfrag{phi_k}{$\phi^v_\gamma$}
\psfrag{particles inside}{$inside \, particle$}
\psfrag{central particle}{$central \, particle, \, r^m_\alpha$}
\centerline{\includegraphics[width=7cm]{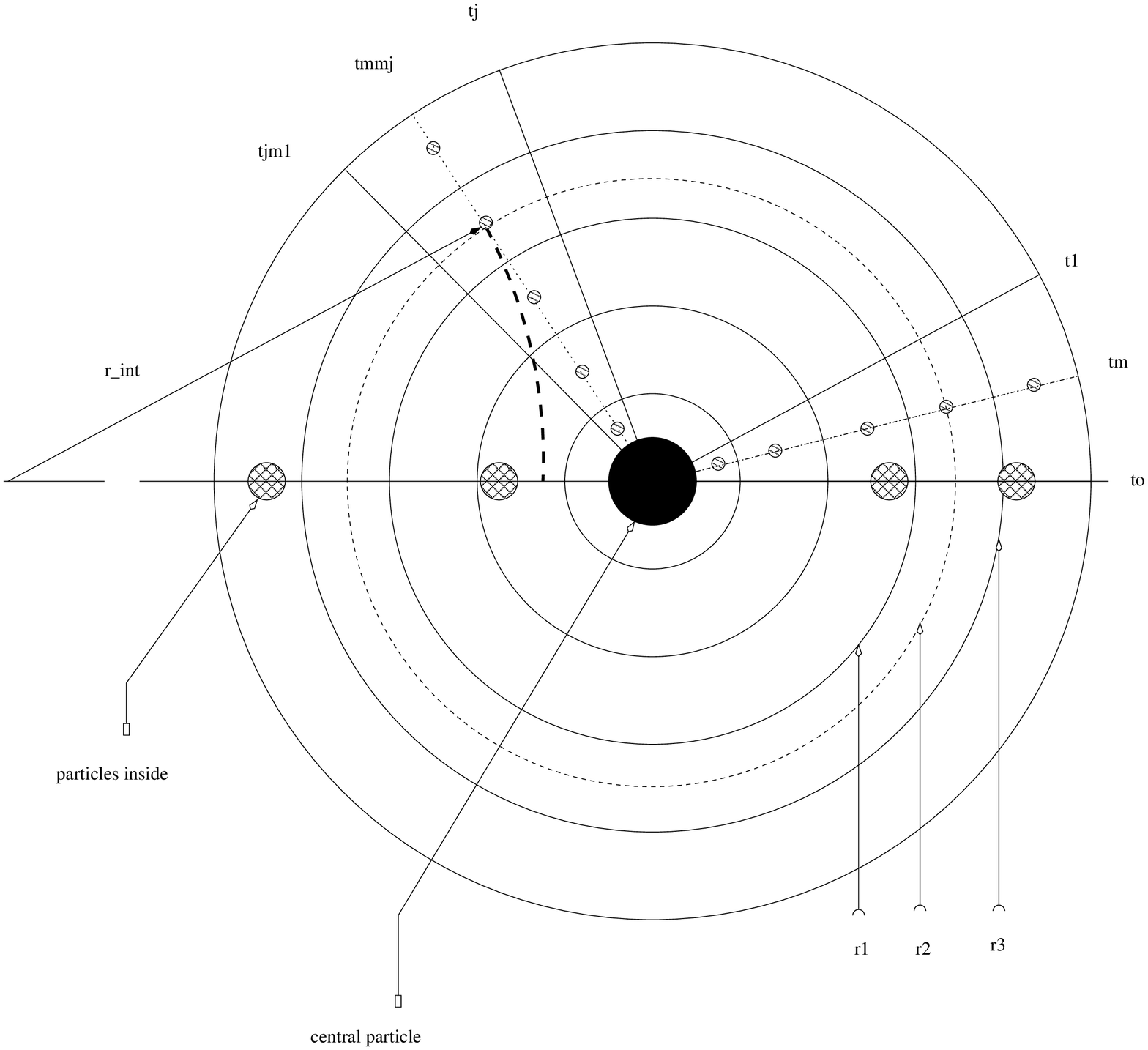}
\includegraphics[width=7cm]{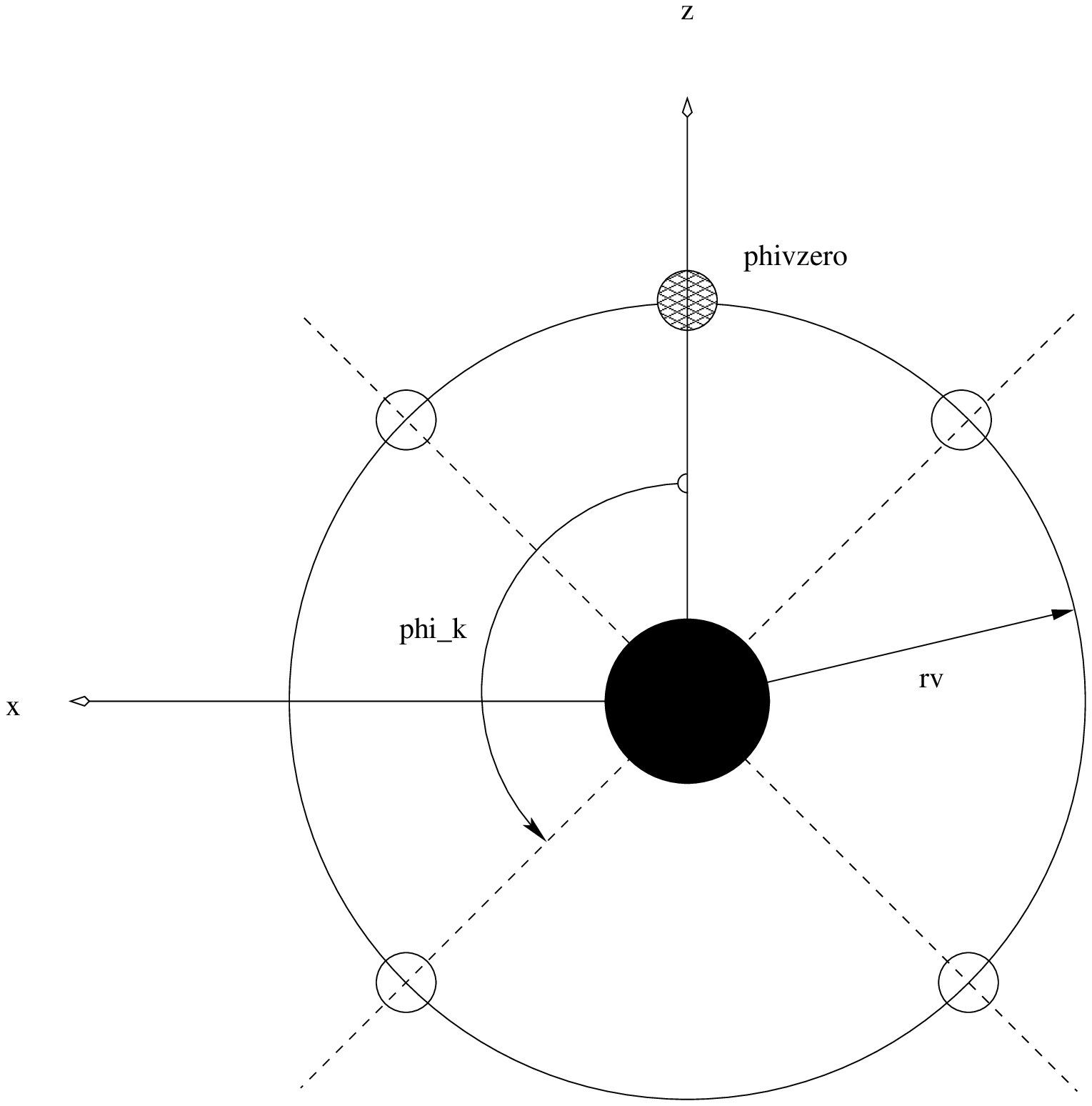}}
\caption{Left. Location of the virtual particles in the $xz$-plane 
($\phi^v=0$). We call this set of points the {\it fan}. 
Given the radius $r^{int}_{\alpha,\beta}$ we interpolate using the real 
particles. 
Right. We show how to assign values to the virtual particles copying
the values of the particles in the fan, due to the one to one relation
between the particles in the fan and the particles at each $\phi^v_\gamma$.}
\label{fig:cartoon2}
\end{figure}

The coordinates assigned to each one of the particles in the {\it fan} are 
given by

\begin{eqnarray}
x^{fan}_{\alpha,\beta}&=&r^v_\alpha \sin \theta^v_\beta \nonumber \\
y^{fan}_{\alpha,\beta}&=&0 \nonumber \\
z^{fan}_{\alpha,\beta}&=&r^m_i + r^v_\alpha \cos\theta^v_\beta .
\end{eqnarray}

The central particle coincides with the physical particle:
\begin{eqnarray}
  \rho^v_{central}&=&\rho_{i}, \nonumber \\
  u^v_{central}&=&u_{i}, \nonumber \\
  v^{v,radial}_{central}&=&v^z_{i},\nonumber \\
  p^v_{central}&=&p_{i},
\end{eqnarray}

and we assign physical values to the virtual particles,
interpolating the values of the real particles using the radial distance
to the origin.

Once the fan has been filled, we can copy for all the virtual
particles in the sphere rotating around the $z$-axis, i.e., for all the
$\phi^v_\gamma$, see Figure \ref{fig:cartoon2}.

For the velocity, we interpolate the radial velocity
$v^{v,radial}_{\alpha,\beta}$, then we reconstruct the cartesian
components of the velocity such that we can introduce them 
in the SPH 3D algorithm. We use the unitary vector 
pointing from the origin to the particle
\begin{equation}
\vec{v}^v_{\alpha,\beta,\gamma}=v^{v,radial}_{\alpha,\beta} \hat{e}_{\alpha,\beta,\gamma} 
\end{equation}
with
\begin{equation}
\hat{e}_{\alpha,\beta,\gamma}=\frac{1}{r^{int}_{\alpha,\beta}}\left(
x^v_{\alpha,\beta,\gamma},y^v_{\alpha,\beta,\gamma},z^v_{\alpha,\beta,\gamma} 
\right)
\end{equation}
and 
\begin{eqnarray}
x^v_{\alpha,\beta,\gamma}&=&r^v_\alpha \sin{ \theta^v_\beta} \cos{\phi^v_\gamma}, 
\nonumber \\
y^v_{\alpha,\beta,\gamma}&=&r^v_\alpha \sin{ \theta^v_\beta} \sin \phi^v_\gamma, \\
z^v_{\alpha,\beta,\gamma}&=&r^m_i+r^v_\alpha \cos{ \theta^v_\beta}. \nonumber
\end{eqnarray}

We identify each $(\alpha,\beta,\gamma)$ with a virtual particle and
the implementation in the 3D code is straightforward.

\subsection{The Virtual Particle Approximation}

\begin{figure}[ht]
\psfrag{Volume errors}{(a)}
\psfrag{title 1y}{$E_V$}
\psfrag{title 1x}{$h$}
\psfrag{Kernel}{(b)}
\psfrag{title 2y}{$E_W$}
\psfrag{title 2x}{$h$}
\psfrag{Kernel Derivative}{(c)}
\psfrag{title 3y}{$E_{DW}$}
\psfrag{title 3x}{$h$}
\psfrag{Rho Approx}{(d)}
\psfrag{title 4y}{$E_\rho$}
\psfrag{title 4x}{$x$}
\centerline{\includegraphics[width=15cm]{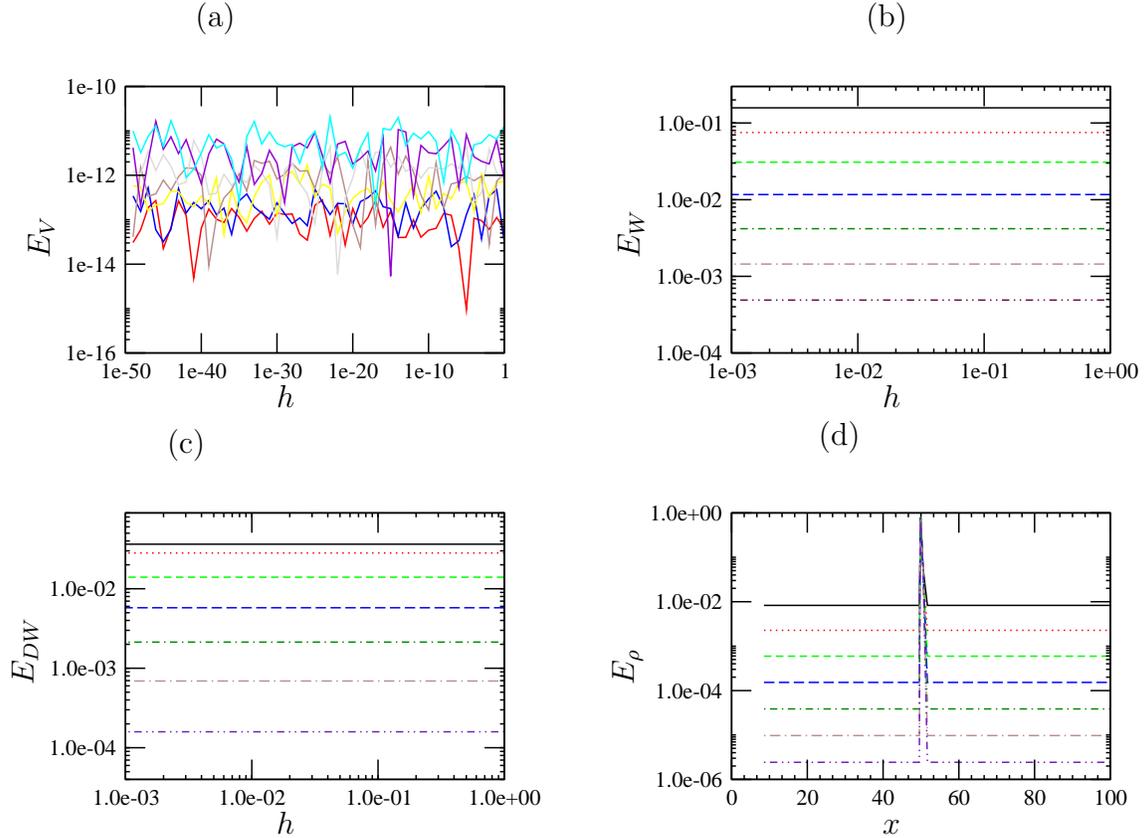}}
\caption{Panel (a) shows the error $E_V$ (defined in the main text) 
as a function of the smoothing length. Each line is obtained with 
a different numbers of virtual particles. The parameters used for
this analysis are $N_v=N_r\cdot N_\theta \cdot N_\phi$, where 
$N_r=2^n \cdot 10$, $N_\theta=80$ and $N_\phi=80$, and $n=0,1,\dots,6$. 
Panel (b) shows the error $E_W$ for the kernel approximation using 
the same number of virtual particles ($N_v$) as before. It is clear 
the convergence of the error to zero. 
Panel (c) shows the error $E_{DW}$ of the derivative of the kernel 
in the $z$ direction. We observe the same behavior as in panel (b). 
Panel (d) presents the error $E{\rho}$ from a discontinuous initial value of 
the density profile. Comparing the exact and the approximated profiles, we 
notice that near the discontinuity, the relative error does not
converge to zero, but in all the other regions it has a clear convergence
for different numbers of virtual particles.}
\label{fig:convergence_virtual}
\end{figure}

Before presenting the simulations obtained using the Cartoon SPH we 
verify that the construction of the virtual particles was made in a
consistent way. We present four convergence tests and we present the
results in Figure \ref{fig:convergence_virtual}

\begin{itemize}

\item {Volume approximation:}

We check that the volume of the virtual particles and the positions assigned 
to them are correct. We compare for a given sphere of radius $h$ its
analytical volume $V_h=\frac{4 \pi }{3} h^3 $ and the approximated 
value of the volume using the volumes of the virtual particles (\ref{eq:v_v})

\begin{equation}
<V>:=\Sigma_j \Delta V_j
\end{equation}

where $\Delta V_j$ is the volume of each one of the $N_v$ particles. 
Figure \ref{fig:convergence_virtual}(a) shows the behavior of the
relative error ($E_v= \frac{|V_h - <V>|}{V}$) as function of the 
smoothing length. We can observe that the relative
error is always close to the round-off error of the computer. 

\item {Normalization of the kernel:}

We verify that the relation $\int W dV=1$ is properly satisfied.
For the i$-th$ physical particle we have
\begin{equation}
<W> := \Sigma_j W_{ij} \Delta V_j \approx 1,
\end{equation}

then, we compute the error $E_W= |1.0-<W>|$ increasing the number 
of virtual particles. We present the behavior in 
Figure \ref{fig:convergence_virtual}(b).
	
\item {Derivative of the kernel:}

We verify that the approximation for the derivative 
$\int (\mbox{\boldmath$r$}-\mbox{\boldmath$r$}') \nabla 
W(\mbox{\boldmath$r$}-\mbox{\boldmath$r$}') d \mbox{\boldmath$r$}' = 1$
is satisfied. The discretization of this equation is
\begin{equation}
<D_iW> := \Sigma_j (\mbox{\boldmath$r$}_i-\mbox{\boldmath$r$}_j) 
\nabla_i W_{ij} \Delta V_j \approx 1 \,\,.
\end{equation}
We compute the relative error $E_{D_iW}:= |1.0-<D_iW>|$ and
obtain proper convergence for all the components of the derivative.
We show the error of the derivative in the $z$ direction in Figure 
(\ref{fig:convergence_virtual}(c)).
	
\item {Density}

The last convergence test compares the numerical approximation of a 
discontinuous profile of density with an initial profile ($\rho_{exact}$)
in a given region of the space:
\begin{equation}
<\rho>_i:= \Sigma_j \rho_j W_{ij} \Delta V_j \,\, ,
\end{equation}
where the subindex $i$ labels the position $\mbox{\boldmath$r$}_i$. 
Again, the relative error is 
$E_{\rho,i} := \frac{|\rho_{exact,i}- <\rho>_i|}{\rho_{exact,i}}$, and the
results are presented in the figure (\ref{fig:convergence_virtual}(d)).
	
\end{itemize}

It is important to notice that if we choose a fixed number of particles
and change the value of the smoothing length, the relative error remains
constant. 

Now we can proceed to present the main simulations used to test the
cartoon implementation.
	
\section{Blast wave}
\label{sec:tests}

We are going to consider a spherical fluid distribution with two regions: 
the first one $r_I \in [0,R_{in}]$ with  $\rho_I=1.0 \times 10^5, 
\epsilon_I=2.5 \times 10^{-5}, v_I=0$ and the second one $r_{II} \in [R_{in}, 
R_T]$ with $\rho_{II}=0.125 \times 10^5, \epsilon_{II}=2.0 \times 10^{-5}, 
v_{II}=0$, here $R_{in}=50$ and $R_T=100$. We are using units where $c=1$.

We assume a flat spacetime properly described by the Minkowsky 
metric $g_{\mu \nu}= diag(-1,1,1,1)$. It is clear that the lapse 
function is $\alpha=1$ and the shift vector $\beta^j=0$, $j=1,2,3$. 
The $3-metric$ induced on the hyper surface of constant time $t$, 
$\Sigma_t$ ,is $\eta_{ij}=\delta_{ij}$ (the Kronecker-delta).

The parameters for the artificial viscosity are 
$\tilde{\alpha}= 1.0$ , $\tilde{\beta}=2.0$ and $\tilde{\epsilon}=0.01$.

To find the initial distribution of the real particles we use equation 
(\ref{eq:find_shells_t})

\begin{equation}
\frac{4 \pi \rho_{in,out}}{3} \left( (r^s_{i})^3-(r^s_{i-1})^3\right)=
\frac{M_{in,out}}{N_s},
\end{equation}
	
with $M_{in,out}$ the mass contained in the inner and outer regions.
In region $I$ the index $i$ takes the values $i=1,\dots,N_{in}$ and in 
region $II$, $i=N_{in}+1,\dots, N_s={N_{in}+N_{out}}$. 
	
We consider $r^s_0=0$ as the origin and $r^s_{N_s}=R_T$ the radius of 
the complete sphere. Then we can get easily the recurrence equation
	
\begin{itemize}
\item {\bf Region I:} 
\begin{equation}
r^s_{i}=\left( \frac{M_{in}}{N_{in}} \frac{3}{4 \pi \rho_I}+(r^s_{i-1})^3 \right)^{1/3},
\end{equation}
			
where $i=1,\dots,N_{in}$.
\item {\bf Region II:}
\begin{equation}
r^s_{i}=\left( \frac{M_{out}}{N_{out}} \frac{3}{4 \pi \rho_{II}}+
(r^s_{i-1})^3 \right)^{1/3},
\end{equation}
			
where $i=N_{in}+1,\dots,N_s$.
\end{itemize}
	
The implementation of equation (\ref{eq:find_shells}) is straightforward
\begin{equation}
r^m_i=\left( \frac{(r^s_{i+1})^3+(r^s_i)^3}{2} \right)^{1/3}. 
\end{equation}
In this test we have used $N_{in}=N_{out}=400$ and $N_s=800$.
	
\subsection{The evolution}
\begin{figure}[ht]
\psfrag{z}{$z$}
\psfrag{Density}{$\rho$}
\psfrag{Energy}{$\epsilon$}
\psfrag{Pressure}{$p$}
\psfrag{Velocity}{$v_z$}
\centerline{\includegraphics[height=11cm]{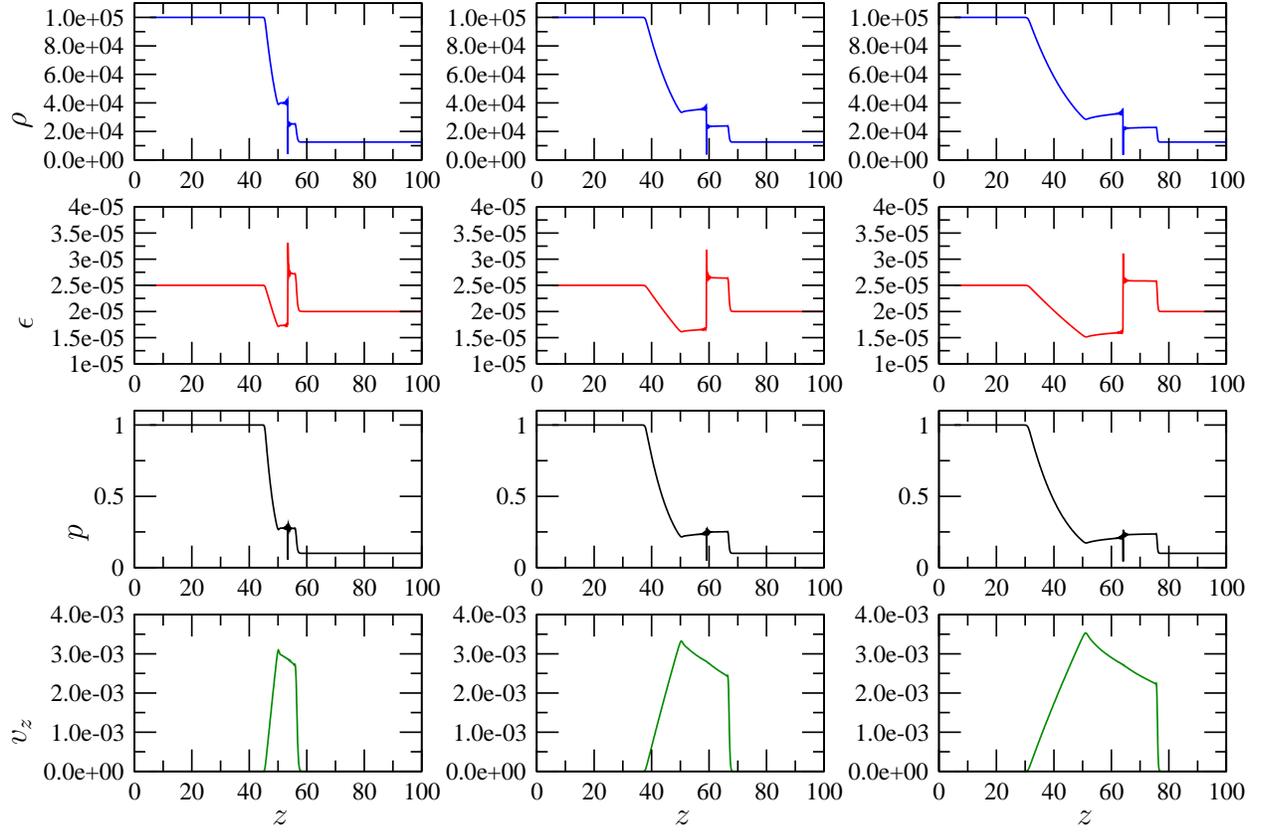}}
\caption{In this figure we present three columns: the first one corresponds 
to time $t=1200$, the second to time $t=3200$ and the third to time $t=5000$. 
The rows are such that the first one is the density profile, the second 
is the specific internal energy, the third is the thermodynamical pressure 
and the fourth one is the velocity in the $z$ direction, which corresponds 
to the radial velocity. All the physical quantities present the same 
structure as in the shock tube (the order of the head, tail, 
contact discontinuity and shock point),  but with differences in the profile.}
\label{fig:evolution}
\end{figure}

We present in the Figure (\ref{fig:evolution}) the physical quantities 
for the blast wave configuration, with the parameters mentioned above. 
The figure contains three different columns corresponding to three different 
times $t=1200, t=3200$ and $t=5000$. In this panels we can appreciate 
the behavior of the physical quantities where the velocities of the
particles are very small compared with the speed of light $(v<<c)$
\cite{Omang}.

We observe (from left to right in Figure \ref{fig:evolution}) 
the existence of the {\it head}, {\it rarefaction wave}, {\it tail},
{\it contact discontinuity} and {\it shock} waves.

The behavior of the simulation is similar to the shock tube with 
some differences in the profile of the velocity between the regions 
of rarefaction and shock. The boundaries of the rarefaction zone are
called the {\it head} and the {\it tail}, \cite{Omang,SieglerRiffert}.

The pressure is continuos in the zone between the tail and 
shock point (in the classical problem this is supported by the 
Rankine-Hugoniot conditions).
	
In the specific internal energy we notice a difference between the 
shock tube and the blast wave, in the region after the rarefaction 
and the contact discontinuity. All these deformations are result of 
the spherical symmetry of the problem.
	
\begin{figure}[ht]
\psfrag{Volume errors}{(a)}
\psfrag{za}{$z\,\,(a)$}
\psfrag{zb}{$z\,\,(b)$}
\psfrag{p}{$p$}
\centerline{\includegraphics[width=14cm]{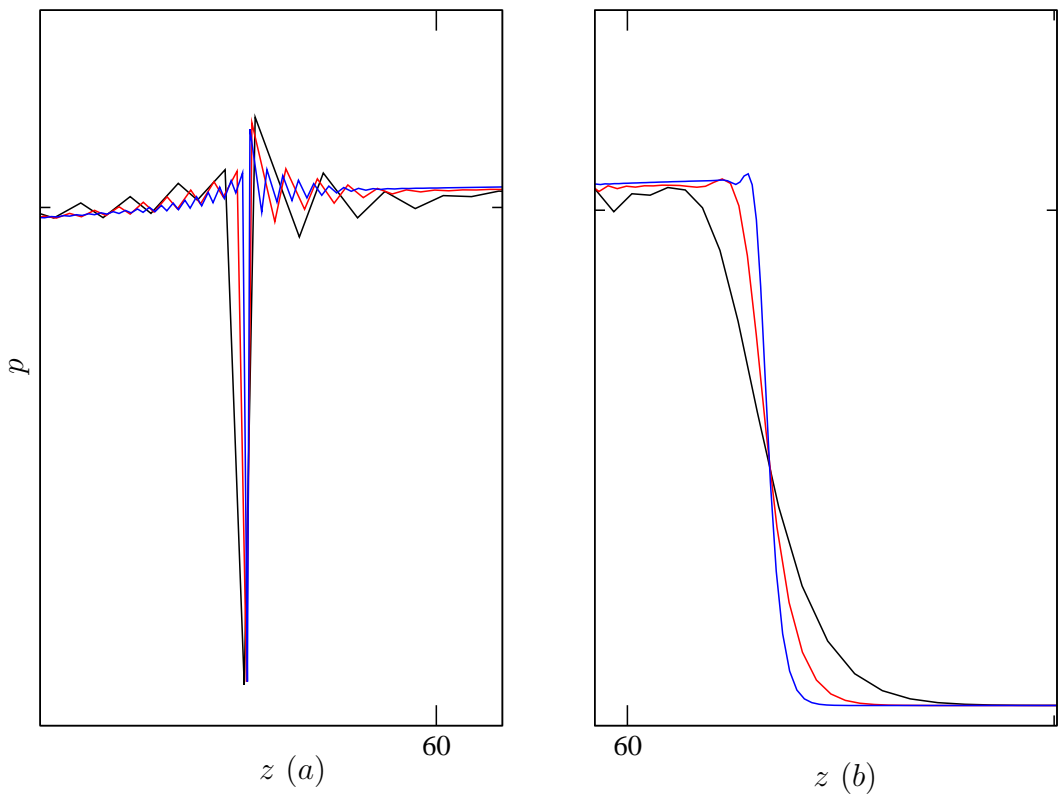}}
\caption{Panel (a) In this panel we present the oscillations around the 
contact discontinuity and how they decrease when we increase the 
number of real particles $N_s=200,400,800$. Panel (b) we can see how 
the numerical solution approximate the step function better if the 
number of particles increase as above.}
\label{fig:oscillations}
\end{figure}

It is clear that all the evolutions have oscillations in the area of the 
contact discontinuity. This is the result of the implementation of the 
artificial viscosity. It is important to mention that if we increase the 
number of real particles in the tests this oscillations decrease as we 
present in Figure (\ref{fig:oscillations}) and it is because 
the linear interpolation is better with more subdivision in the radial 
direction. We used three different values of $N_s=200,400,800$. 
It is also possible to used better interpolations to decrease the 
oscillations. 

\section{Conclusions}
The implementation introduced in this article is a successful method 
that can be used to deal with problems in spherical symmetry with 
3 dimensional codes instead of rewriting the equations in that symmetry. 
This implementation is simple and can be also applied to problems with
axial symmetry. In the same way, it can also be implemented for the
Newtonian Euler equations \cite{Cruz}.
The tests we presented here, give a clear idea of the behavior of the
cartoon and can be used to perform further analysis and studies of
physical and astrophysical scenarios.
\label{sec:conclusions}

\section{Acknowledgments}
This work is supported by grants CIC-UMSNH-4.23,
PROMEP UMICH-CA-22, UMICH-PTC-210 and CONACyT 79601.

\end{document}